\documentstyle[preprint,aps,epsfig]{revtex}
\begin{document}
\draft
\title{
{\rm \today} \\
Electronic structures of La$_3$S$_4$ and Ce$_3$S$_4$}
\author{J. H. Shim$^1$, Kyoo Kim$^1$, B. I. Min$^1$, and J.-S. Kang$^2$}
\address{
$^1$ Department of Physics, Pohang University of Science and
Technology, Pohang 790-784, Korea \\
$^2$Department of Physics, The Catholic University of Korea,
        Puchon 420-743, Korea}
%\date{}
\maketitle
%%%%%%%%%%%%%%%%%%%%%%
\begin{abstract}
We have investigated electronic structures of La$_3$S$_4$ and Ce$_3$S$_4$
using the LSDA and LSDA+$U$ methods. Calculated density of states (DOS) 
are compared with the experimental DOS obtained by the valence band
photoemission spectroscopy. The DOS at $E_{\rm{F}}$ 
indicates the 5$d$ character in La$_3$S$_4$ and 4$f$ character in Ce$_3$S$_4$.
It is found to be nearly half metallic in the ferromagnetic ground state of
Ce$_3$S$_4$.
%Ce$_3$S$_4$ has ferromagnetic ground states with spin and orbital magnetic
%moments of 1.27 $\mu_{\rm{B}}$ and $-$2.81 $\mu_{\rm{B}}$ per Ce, respectively,
%and shows nearly half metallic ground state. 
We discuss the superconductivity and structural transition in
La$_3$S$_4$, and the absence of structural transition in Ce$_3$S$_4$.
\end{abstract}
\pacs{Keywords: La$_3$S$_4$, Ce$_3$S$_4$, electronic structure, band Jahn-Teller
\\
\begin{quote}
Corresponding author:\\
               Prof. B. I. Min,\\
               Department of Physics,\\
               Pohang University of Science and Technology,\\
               Pohang, 790-784, KOREA\\
               (e-mail) bimin@postech.ac.kr\\
               (Fax) +82-54-279-3099\\
\end{quote}
}
%
%\narrowtext
%
The Th$_3$P$_4$-type La$_3$S$_4$ undergoes cubic to tetragonal transition at
$\sim$ 103 K, and shows superconductivity at $T_c$ = 8.3 K\cite{wester79}.
%La$_{3-x}$S$_4$ has the single phase of Th$_3$P$_4$ structure upto 
%$x = \frac{1}{3}$, where it changes metallic to semiconducting.
%Increase of vacancy concentration $x$ on the La sites lowers the structural
%transition temperature\cite{wester80}. Less than 1\% of vacancies on the La 
%site suppresses the tetragonal phase transition completely. 
%It thus suggests that 
%the structural phase transition depends on the conduction-electron 
%concentration. 
On the other hand, Ce$_3$S$_4$ with the same structure shows
no tetragonal distortion and has ferromagnetic ground state with 
the transition temperature of $\sim$ 7.2 K\cite{futt88}.
The structural transition in La$_3$S$_4$ has been explained by the band 
Jahn-Teller mechanism, in which the degenerate La $e_g$ band induces the
tetragonal phase transition\cite{ray81,kim88}. 
It has been suggested that the existence of 4$f$ band at Fermi level
($E_{\rm F}$) in Ce$_3$S$_4$ prevents the band Jahn-Teller type structural
transition\cite{cho00}. However the electronic structures of La$_3$S$_4$
and Ce$_3$S$_4$ have not been confirmed yet.

In this study, we have investigated electronic structures of La$_3$S$_4$
and Ce$_3$S$_4$ using the local spin-density approximation (LSDA) and 
the LSDA+$U$
%(LSDA incorporating the on-site Coulomb interaction $U$) 
method on the basis
of the linearized muffin-tin orbital (LMTO) band method.
In the LSDA+$U$ method, the spin-orbit interaction is taken into
account in the
self-consistent variational loop, so that the orbital polarization is
properly taken into account\cite{kwon00}.
The angular momentum projected local density of states (PLDOS)
is compared to the result of photoemission spectroscopy (PES). 
%PES experiments were carried out at the Synchrotron Radiation Center (SCR).
Samples were
fractured and measured in vacuum with a base pressure better than 
$3\times10^{-11}$ Torr and at T$\approx$15 K. Experimental details will be
reported elsewhere\cite{kang02}.

%The electronic structure of La$_3$S$_4$ are calculated for paramagnetic ground
%state in the LSDA. 
The LSDA DOSs of La$_3$S$_4$ in the paramagnetic ground
state are shown in Fig.~\ref{ls}. 
La 5$d$ and S 3$p$ PLDOSs are very similar each other, reflecting the large
hybridization between them. 
%S 3$p$ states are almost occupied at
%$-$3 eV $\sim$ $-$7 eV. 
Small DOS peak appears near $E_{\rm F}$ corresponding
to the tail of huge unoccupied DOS above $E_{\rm F}$. The main contribution to
$N(E_{\rm F})$ comes from La 5$d$ states ($\sim$ 60\%). The superconductivity
of La$_3$S$_4$ is due to the high DOS of La 5$d$ states at $E_{\rm F}$.
From the cubic to tetragonal structure, the bands at $E_{\rm F}$ are split
and the DOS becomes a bit wider, as is consistent with the band Jahn-Teller 
model.
%But the width of La band near $E_{\rm F}$ ($\sim$ 200 meV) is larger by almost
%one order than the fitting parameter employed in the analysis of 
%the band Jahn-Teller model\cite{cho00}. 
But
Fig.~\ref{ls}(b) shows that $t_{2g}$ states are dominant at $N(E_{\rm F})$
rather than the $e_g$ states. 
Hence, it is not compatible with the band Jahn-Teller model 
assuming the degenerate $e_g$ bands near $N(E_{\rm F})$. 
The present study indicates that a modified band Jahn-Teller model
considering realistic band structures is required.

Figure~\ref{cs}(a) provides the DOSs of ferromagnetic Ce$_3$S$_4$ 
obtained by using the LSDA+$U$ method. 
We use the parameters of the on-site Coulomb interaction $U$ = 7.0 eV
and the exchange $J$ = 0.98 eV for the Ce 4$f$ electrons. 
%The effect of on-site
%Coulomb interaction makes the majority occupied spin 4$f$ band moves down to
%higher binding energy $-$2 eV, which is originally concentrated around
%$E_{\rm F}$ in the LSDA. 
The majority occupied spin 4$f$ band is located near $-$2 eV, and
the second majority spin 4$f$ band is near 
$E_{\rm F}$, whereas minority spin 4$f$ bands are all above $E_{\rm F}$.
The minority spin states are negligible at $E_{\rm F}$, suggesting that
Ce$_3$S$_4$ is nearly half metallic in the ferromagnetic ground state. 
The large contribution
to $N(E_{\rm F}$) comes from mainly Ce 4$f$ states ($\sim$87\%), which is 
compatible with the previous analysis that the existence of $f$ electrons 
at $E_{\rm F}$ prevents the
band Jahn-Teller type structural transition in Ce$_3$S$_4$\cite{cho00}.
The calculated spin and orbital magnetic moments for Ce are 1.27 $\mu_{\rm B}$
and $-$2.81 $\mu_{\rm B}$, respectively.

Figure~\ref{cs}(b)-(d) compare the PLDOSs to the measured partial
spectral weight (PSW).
Figure~\ref{cs}(b) presents the comparison of Ce 4$f$ PLDOS with the
extracted Ce 4$f$ PSW obtained by
the Ce 4$d \rightarrow$ 4$f$ on-resonacne spectrum ($h\nu$ = 121 eV).
Ce 4$f$ PSW exhibits the prominent features at $-$0.5 eV, $-$2.5 eV, and
$-$5.5 eV. Although the main Ce 4$f$ PLDOS states are $\sim 0.5$ eV off 
from the experimental result, the LSDA+$U$ 
calculation provides a reasonably good description of electronic structure
of Ce$_3$S$_4$. Likewise, there are good agreements between
Ce 5$d$ and S 3$p$ PLDOSs and the Ce 5$d$ ($h\nu$ = 26 eV) and
S 3$p$ ($h\nu$ = 18 eV) PSWs.

In conclusion, we have studied the electronic structures of La$_3$S$_4$
and Ce$_3$S$_4$ using the LSDA and LSDA+$U$ methods, and the PLDOSs are
compared to the measured PSWs. 
%Both compounds show the strong hybridization
%between La/Ce 5$d$ and S 3$p$ states. 
The DOS of La$_3$S$_4$ shows the
high contribution from the La 5$d$ character at $E_{\rm F}$.
For the ferromagnetic Ce$_3$S$_4$, the DOS at $E_{\rm F}$ shows mainly 
Ce 4$f$ character.
% which is consistent with the 
%previous suggestion that the structural transition is suppressed by Ce 4$f$ 
%states at $E_{\rm F}$. 
%It is found to be nearly half metallic.
Nearly half metallic LSDA+$U$ band structure for Ce$_3$S$_4$ provides 
the good agreement with experimental results.

%\acknowledgments
Acknowledgments$-$
This work was supported by the KRF grant (KRF-2002-070-C00038).
JSK acknowledges the support from the Catholic University of Korea (20020019).

%%%%%%%%%%%%%%%%%%%%%%%%%%%%%%

\begin{figure}
%\centerline{\epsfig{figure=fig1.ps,angle=270,width=7.5cm}}
\caption{
(a) Total DOS of La$_3$S$_4$ in the paramagnetic ground state.
(b) La 5$d$ ($t_{2g}$ for dotted line) and (c) S 3$p$ PLDOSs.
%Total and projected local density of states of La$_3$S$_4$ in the paramagnetic
%ground state.
    }
\label{ls}
\end{figure}
\begin{figure}
%\centerline{\epsfig{figure=fig2.ps,angle=270,width=7.5cm}}
\caption{
(a) Spin polarized total DOS of Ce$_3$S$_4$ in the ferromagnetic ground state.
(b)-(d) Ce 4$f$, Ce $5d$, and S 3$p$ PLDOSs are compared to the experimental
Ce 4$f$, the $h\nu$ =26 eV, and $h\nu$ = 18 eV PSWs (dotted lines).
The majority and monority spin DOSs are summed for comparison. 
    }
\label{cs}
\end{figure}

\newpage
\centerline{Fig. 1.}
\centerline{\epsfig{figure=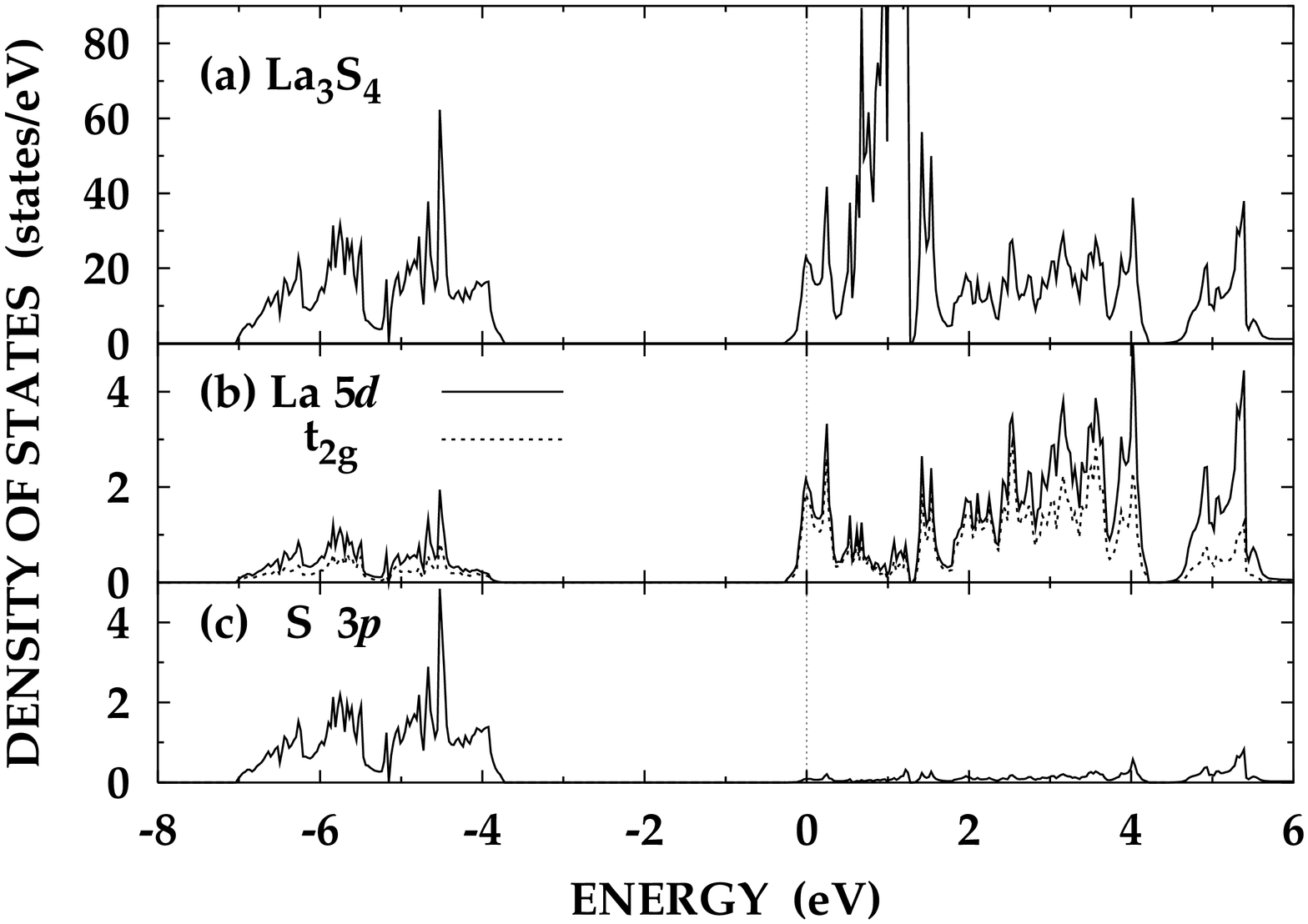,angle=270,width=14cm}}

\centerline{Fig. 2.}
\centerline{\epsfig{figure=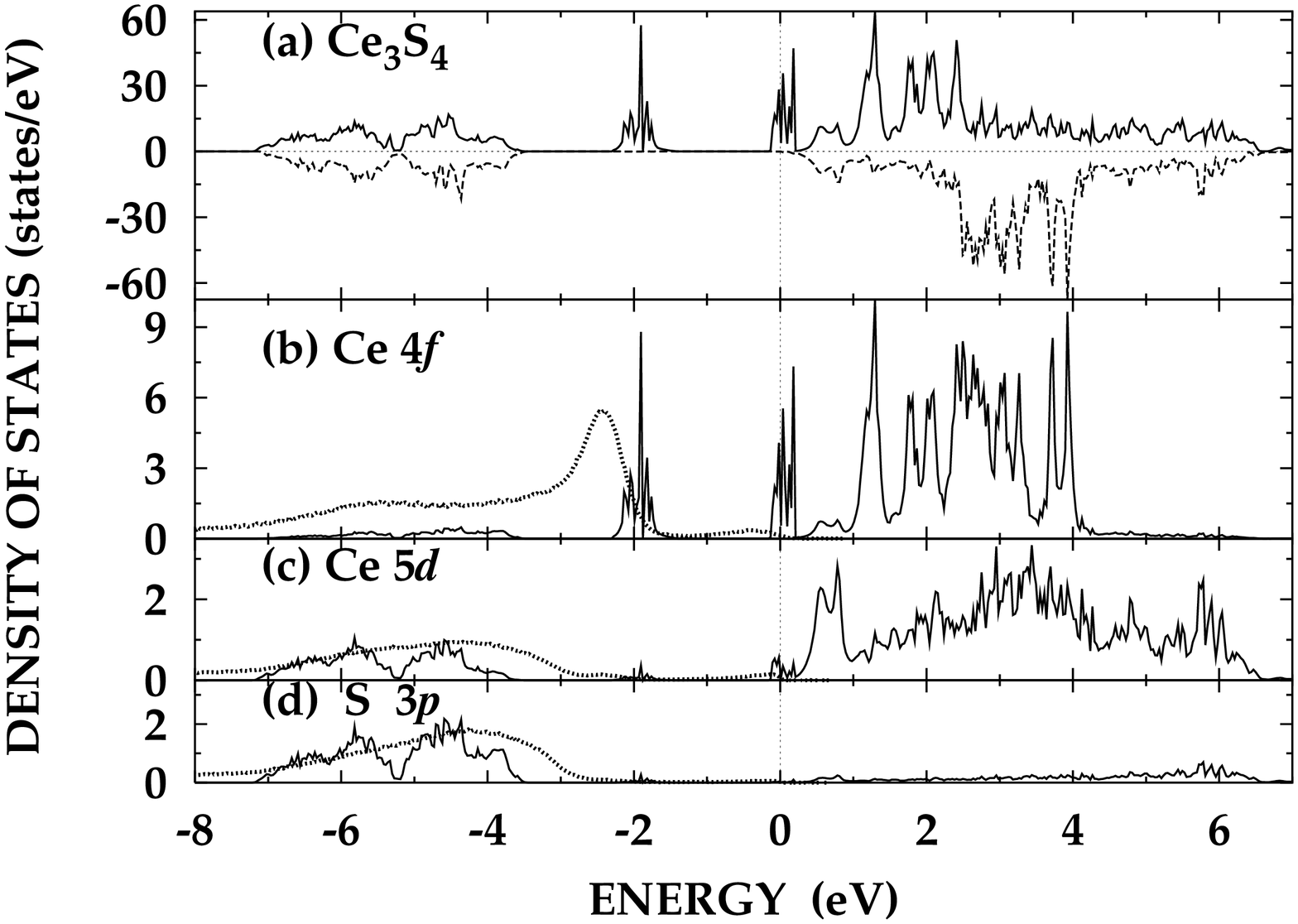,angle=270,width=14cm}}

\end{document}